\pdfoutput=1
\documentclass[preprint,superscriptaddress,longbibliography]{revtex4-2}
\usepackage{graphicx}
\usepackage{dcolumn}
\usepackage{bm}
\usepackage{epstopdf}
\usepackage{xr}
\usepackage{amsmath}
\usepackage{color}
\usepackage[normalem]{ulem}
\usepackage{wasysym}
\usepackage{siunitx}
\usepackage{physics}
\usepackage{lineno}
\usepackage[T1]{fontenc}
\usepackage{booktabs}
\usepackage[section]{placeins}
\usepackage{textcomp}
\usepackage{amssymb}
\usepackage{xcolor}
\usepackage[explicit]{titlesec}
\usepackage{multibib}

\DeclareUnicodeCharacter{0306}{ }
\DeclareUnicodeCharacter{03B4}{ }
\DeclareUnicodeCharacter{0394}{$\Delta$}
\DeclareUnicodeCharacter{2212}{\textendash}
\UseRawInputEncoding

\usepackage{hyperref}
\hypersetup{colorlinks=true,%
citecolor=blue,%
filecolor=black,%
linkcolor=black,%
urlcolor=blue
}

\begin{document}

\makeatletter
\newcommand{\rmnum}[1]{\romannumeral #1}
\newcommand{\Rmnum}[1]{\expandafter\@slowromancap\romannumeral #1@}
\makeatother

\renewcommand{\figurename}{Figure}

\title{Nanowire bolometer using a 2D high-temperature superconductor}

\author{Sanat Ghosh}
\homepage{sanatghosh1996@gmail.com}
\affiliation{Department of Condensed Matter Physics and Materials Science, Tata Institute of Fundamental Research, Homi Bhabha Road, Mumbai 400005, India}

\author{Digambar A. Jangade}
\affiliation{Department of Condensed Matter Physics and Materials Science, Tata Institute of Fundamental Research, Homi Bhabha Road, Mumbai 400005, India}

\author{Mandar M. Deshmukh}
\homepage{deshmukh@tifr.res.in}
\affiliation{Department of Condensed Matter Physics and Materials Science, Tata Institute of Fundamental Research, Homi Bhabha Road, Mumbai 400005, India}

\begin{abstract}

Superconducting nanowires are very important due to their applications ranging from quantum technology to astronomy. In this work, we implement a non-invasive process to fabricate nanowires of high-$T_\text{c}$ superconductor Bi$_2$Sr$_2$CaCu$_2$O$_{8+\delta}$ (BSCCO). We demonstrate that our nanowires can be used as bolometers in the visible range with very high responsivity of 9.7 $\times$ 10$^{3}$ V/W. Interestingly, in a long (\SI{30}{\micro\meter}) nanowire of 9~nm thickness and 700~nm width, we observe bias current-dependent localized spots of maximum photovoltage. Moreover, the scalability of the bolometer responsivity with the normal state resistance of the nanowire could allow further performance improvement by increasing the nanowire length in a meander geometry. We observe phase slip events in nanowires with small cross-sections (12~nm thick, 300~nm wide, and \SI{3}{\micro\meter} long) at low temperatures. Our study presents a scalable method for realizing sensitive bolometers working near the liquid-nitrogen temperature. 

\end{abstract}


\maketitle

%-------------------------------------------------------------------

\section{Introduction}

%-------------------------------------------------------------------

Superconducting nanowires have attracted increasing interest in recent years because of their technological applications, as well as for addressing fundamental questions of superconductivity in reduced dimensions. Because of the ultra-small form factor, currently, they are one of the most sensitive photon detectors \citep{hadfield_single-photon_2009}, reaching single photon detection limit \citep{goltsman_picosecond_2001}, having large bandwidth \cite{zinoni_erratum_2010}, low dark count \citep{shibata_ultimate_2015}, and limited dead time \citep{tarkhov_ultrafast_2008}. They have been realized as a nonlinear inductor \citep{ku_superconducting_2010} for a prototypical qubit system in the field of quantum information and technology. Using superconducting nanowire detectors, significant advancements in the field of astronomy have also been possible \citep{hochberg_detecting_2019}. 

%--------------------------------------------------------------------

%--------------------------------------------------------------------

In spite of superior qualities, the applicability of superconducting detectors is limited as they require very low temperature for operation, usually liquid helium temperature. A superconducting detector working above the boiling point of liquid nitrogen, therefore, has a distinct advantage. High-$T_\text{c}$ cuprate superconductors with a particular interest in exfoliable van der Waals material because of their high quality could pave the way in that direction.
%--------------------------------------------------------------------

%-------------------------------------------------------------------

For a long time, superconductivity in layered cuprate superconductors, where superconducting CuO$_2$ planes are separated by insulating charge reservoir layers, was considered a bulk phenomenon. Recent developments in the exfoliation of van der Waals materials, however, show that the superconductivity is achievable down to 0.5 unit cell thick ($\sim$ 1.5 nm) Bi$_2$Sr$_2$CaCu$_2$O$_{8+\delta}$ (BSCCO)--a high-$T_\text{c}$ superconducting material while maintaining $T_\text{c}$ similar to its bulk counterpart \cite{yu_high-temperature_2019,sterpetti_comprehensive_2017}. This suggests superconductivity in cuprate is of 2D nature, entirely arising from single CuO$_2$ planes. Isolating few unit cells thick BSCCO and the ability to modify superconductivity locally can, in principle, allow us to realize nanowire of extremely small form factor. The small form factor of the nanowire is desirable as it has low heat capacity and low thermal conductance (because of the small contact area between bolometer and substrate), which is expected to result in both a fast and a sensitive bolometer \citep{hu_design_1989}. 

%--------------------------------------------------------------------

%--------------------------------------------------------------------

\begin{figure*}
\includegraphics[width=15.5cm]{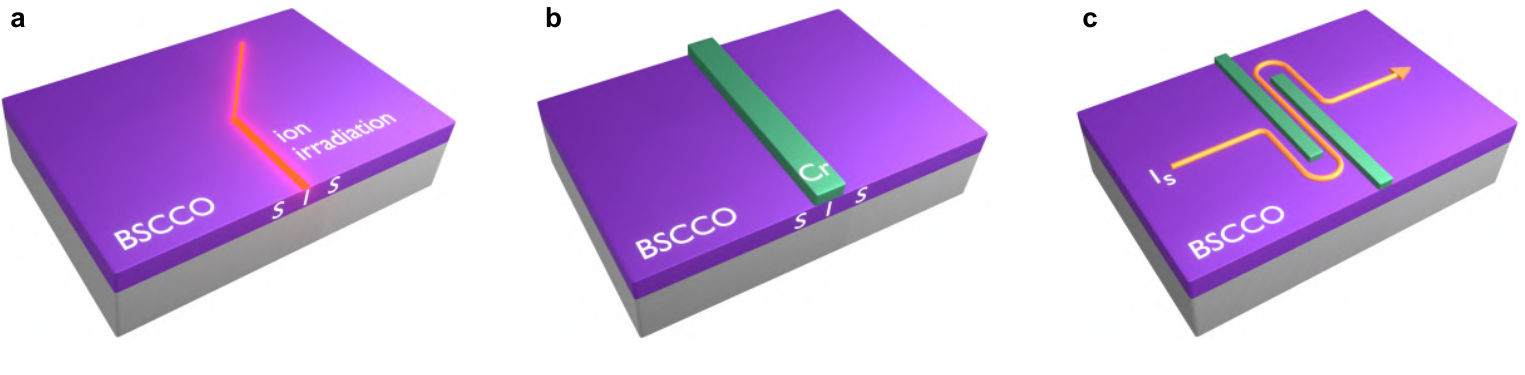}
\caption{ \label{fig:fig1} \textbf{Alternative approach in patterning superconducting nanowires compared to existing method.} (a) Schematic of conventional technique of patterning superconductivity. Exposure to high intensity ion beam locally damages superconducting property making it an insulator. (b) Schematic of our approach to pattern superconductivity locally in BSCCO. Deposition of Cr on selected region makes underneath BSCCO insulating. (c) Patterning BSCCO in nanowire geometry. Two separated Cr lines are deposited on BSCCO which defines the nanowire. Arrow shows the supercurrent ($I_\text{s}$) path through the device.}
\end{figure*}

However, fabricating nanowires using BSCCO is challenging due to their extreme chemical sensitivity. For patterning narrow structures on a planar surface, ion milling or selective etching has been a standard route \citep{curtz_patterning_2010}. A similar approach has been taken where a well-focused high-intensity ion beam is irradiated on a selected region of a thin superconductor to modify its superconductivity \citep{cybart_nano_2015}; this locally damages the superconductor making it an insulator, as shown schematically in Figure~\ref{fig:fig1}a. The above-mentioned conventional methods of patterning superconductivity by locally irradiating high energetic ion beam cause inhomogeneity and disorder in the nanowire, which can compromise the detector's performance. Thus there is a need for an alternative approach to circumvent this issue.

In this letter, we integrate the two aspects -- an alternative scheme for patterning superconducting nanostructure and the use of a few unit cells thick high-$T_\text{c}$ van der Waals superconductor to fabricate nanowires of high-$T_\text{c}$ cuprate superconductor BSCCO. An excellent strategy to realize a high coupling efficiency detector involving superconducting nanowires is to make a long meander geometry \citep{natarajan_superconducting_2012}. To this end, as a proof of concept, we study two sets of nanowire devices with different aspect ratios -- a long nanowire (NW1) with \SI{30}{\micro\metre} length, 700~nm width, and 9~nm thickness and a short nanowire (NW2) with \SI{3}{\micro\metre} length, 300~nm width, and 12~nm thickness. We perform detailed transport characterization and photoresponse of the nanowires and show these nanowires can be used as a sensitive bolometer for visible range. The Long nanowire (NW1) shows localized region of maximum photovoltage, which changes its location along the nanowire length upon changing the bias current. We observe that the responsivity of the superconducting bolometer scales with the normal state resistance of the nanowire. Furthermore, our measurements on short nanowire (NW2) with smaller cross-section suggest the presence of phase slips.

%--------------------------------------------------------------------

%--------------------------------------------------------------------

%-------------------------------------------------------------------

\section{Method}

%------------------------------------------------------------------

As shown in our earlier study \citep{ghosh_-demand_2020}, we selectively deposit Cr on the surface of freshly cleaved few unit cells thick flake of BSCCO, thereby making underneath superconductor insulating, schematically shown in Figure~\ref{fig:fig1}b. Following this strategy, we deposit two Cr lines on exfoliated thin BSCCO with a gap in between the Cr lines to define the nanowire, as shown in Figure~\ref{fig:fig1}c. A capping layer of Au was deposited on the Cr lines to protect the Cr lines from being oxidized. 

Working with very thin layers of BSCCO is challenging as the dopant oxygen atoms diffuse out of the material \citep{poccia_evolution_2011} over time, eventually making it insulating \citep{novoselov_two-dimensional_2005,sandilands_origin_2014}. Various capping layers have been used to minimize this effect in past studies \citep{jiang_high-_2014,zhao_sign-reversing_2019}. Exposure to chemicals while making electrical contacts by standard lithographic process degrades the device quality \citep{vasquez_intrinsic_1994}. To minimize device fabrication time and to avoid chemical treatments, we directly deposit 70~nm thick gold contacts to the freshly exfoliated BSCCO flake through a pre-aligned SiN mask \citep{deshmukh_nanofabrication_1999}. Through a second SiN mask, we then deposit two 15~nm/10~nm of Cr/Au lines to define nanowire geometry, as shown schematically in Figure~\ref{fig:fig1}c. Anodic bonding method \citep{sterpetti_comprehensive_2017} has been used to exfoliate large area ($\sim$ \SI{100}{\micro\metre} lateral dimension) flakes from optimally doped bulk Bi-2212 crystal. After the entire fabrication process, we immediately transfer the device to a high vacuum closed-cycle cryostat. Electrical measurements are done using low-frequency \textit{ac} lock-in detection technique in four-probe configuration. Optical micrographs and SEM images of the nanowires are shown in Supporting Information (Figure~S1).

%-----------------------------------------------------------------

\begin{figure*}
\includegraphics[width=15.5cm]{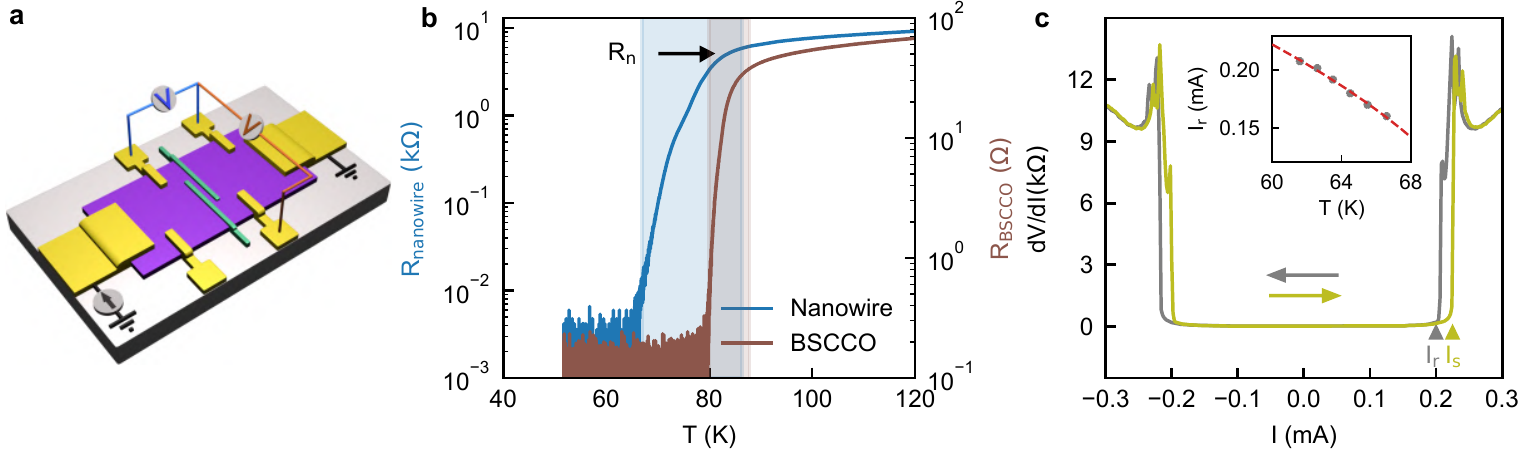}
\caption{ \label{fig:fig2} \textbf{Superconducting nanowire device architecture and its electrical characterization.} (a) Schematic of the nanowire device and the measurement scheme. The electrodes are designed in such a way that we can simultaneously measure response of the nanowire (by blue coloured voltmeter) and the remaining extended BSCCO (by brown coloured voltmeter). This serves as an inbuilt check on quality of the pristine BSCCO from which nanowire is patterned. (b) Simultaneous measurement of four probe resistances versus temperature of the device. Different coloured curves correspond to the response measured by the same coloured voltmeter indicated in (a). The shaded regions indicate width of the superconducting transitions of pristine BSCCO (brown) and the nanowire (blue). We identify $T_\text{c}$ of the pristine BSCCO as the temperature where $dR/dT$ is maximum. Electrode pair across nanowire involve contribution from some part of the extended flake as well. Resistance at $T_\text{c}$ of BSCCO is thus taken as the normal state resistance $R_\text{n}$ of the nanowire. The data corresponds to a 3 unit cells (9~nm) thick BSCCO. (c) Differential resistance ($dV/dI$) vs \textit{dc} biasing current of the nanowire. Two curves are for two sweep directions of biasing current. Retrapping ($I_\text{r}$) and switching ($I_\text{s}$) currents are identified as discussed in the text. The inset shows evolution of $I_\text{r}$ with temperature and its fit with hotspot model (red dashed curve). }
\end{figure*}

%------------------------------------------------------------------

\section{Results and discussion}

We now describe the electrical characterization of the nanowire devices -- first using low biasing current and later at higher biasing current. Figure~\ref{fig:fig2}a schematically shows the transport measurement scheme. We always simultaneously measure response across the nanowire and pristine BSCCO to make sure the parent BSCCO is of good quality and did not degrade during the fabrication steps; this serves as an inbuilt control. Figure~\ref{fig:fig2}b shows the temperature variation of the resistance ($R$ vs. $T$) of NW1, which is fabricated from 3 unit cells (9~nm) thick BSCCO. We define superconducting transition temperature $T_\text{c}$ as the temperature where $dR/dT$ is maximum; using this definition, $T_\text{c}$ across pristine BSCCO is 83~K. The nature of $R$ vs. $T$ across nanowire and pristine BSCCO is similar with a qualitative difference -- superconducting transition across nanowire is broader and has a longer tail compared to the pristine region before it reaches the zero resistance state. The superconducting onset temperatures of pristine BSCCO and nanowire are similar but has transition width of 7.8~K and 19.4~K, respectively, as indicated by the shaded region in Figure~\ref{fig:fig2}b. Due to our device geometry, the electrode pairs that measure response across nanowire also involve a certain part of the extended pristine BSCCO. We thus define the normal state resistance $R_\text{n}$ of the nanowire as the resistance at $T_\text{c}$ of pristine BSCCO. With this definition, $R_\text{n}$ of NW1 is $\sim$ 5.1 k$\Omega$, as indicated by the arrow in Figure~\ref{fig:fig2}b. $R$ vs. $T$ data of different nanowires showing superconducting to insulating transition with reducing nanowire width is shown in Supporting Information (Figure~S2).

%-------------------------------------------------------------------

%-------------------------------------------------------------------

Next, we discuss the response of the long nanowire at a finite biasing current. Figure~\ref{fig:fig2}c shows a plot of differential resistance ($dV/dI$) of the nanowire with \textit{dc} biasing current ($I-dV/dI$ characteristic) at 61.5~K. The response of pristine BSCCO from the same device at different temperatures is shown in Supporting Information (Figure~S3). The two curves are for two sweep directions of the \textit{dc} biasing current indicated by the two arrows, which show hysteresis. While going from zero to a finite resistance state, nanowire switches from superconducting to normal state (S-N transition) at a current referred to as switching current ($I_\text{s}$). While traversing back, the system returns to the superconducting state at retrapping current ($I_\text{r}$), as indicated by the arrowheads in Figure~\ref{fig:fig2}c. The different values of $I_\text{s}$ and $I_\text{r}$, or equivalently the hysteresis in the response of these nanostructures, are generally attributed to heating effects described by the hotspot model \citep{skocpol_selfheating_1974}, where a localized normal region is maintained by Joule heating. The generated heat and its dissipation lead to a characteristic thermal healing length $\eta = \sqrt{\kappa d /\alpha}$ in the system. Here $\kappa$, $\alpha$, and $d$ are thermal conductivity of BSCCO ($\sim$ 2.5 W m$^{-1}$ K$^{-1}$ \citep{yang_thermal_1995}), heat transfer coefficient of BSCCO/glass interface, and thickness of BSCCO ($\sim$ 9 nm), respectively. The value of $\alpha$ for the BSCCO/glass interface is not known. As a reference, we take the value of $\alpha$ to be $\sim$ 850 W cm$^{-2}$ K$^{-1}$ (for YBCO/SrTiO3 interface \citep{nahum_thermal_1991}). For our nanowire length $L$ ($\sim$ \SI{30}{\micro\metre}) much larger than $\eta$ ($\approx$ 50 nm, estimated with quoted values of the parameters), $I_\text{r}$ within the hotspot model is given by \citep{skocpol_selfheating_1974} $I_{r} \approx (2\alpha W^{2}dT_{c} / \rho_{n})^{1/2} (1-T/T_{c})^{1/2}$, where $\rho_{n}$ and $W$, are normal state resistivity and width of the nanowire. The inset of Figure~\ref{fig:fig2}c shows the variation of $I_\text{r}$ with temperature and its fit (red dashed line) using the hotspot model. The fit suggests a good agreement between experimental data and the hotspot model providing values of fitting parameters $T_\text{c}$ and $\alpha$ to be 73.4~K (close to the value of 72.5~K that we get from $R$ vs. $T$) and 45 W cm$^{-2}$ K$^{-1}$. The lower value of $\alpha$, estimated from the fit, suggests that heat transport across BSCCO/glass interface is low. This also means that the thermal healing length $\eta$ will be larger ($\approx$ 224 nm). We studied hysteresis in the nanowire with different shunting resistors. Apart from reducing hysteresis, the shunting resistor does not alter the response of the nanowires significantly. In the main text, we present all the data without a shunt resistor and have shown data with different shunting resistors in Supporting Information (Figure~S5). Apart from hysteresis, we also see multiple resistive jumps in the S-N transition in terms of multiple peaks in $dV/dI$ (as shown in Figure~\ref{fig:fig2}c) that we discuss next.

%-------------------------------------------------------------------

\begin{figure*}
\includegraphics[width=15.5cm]{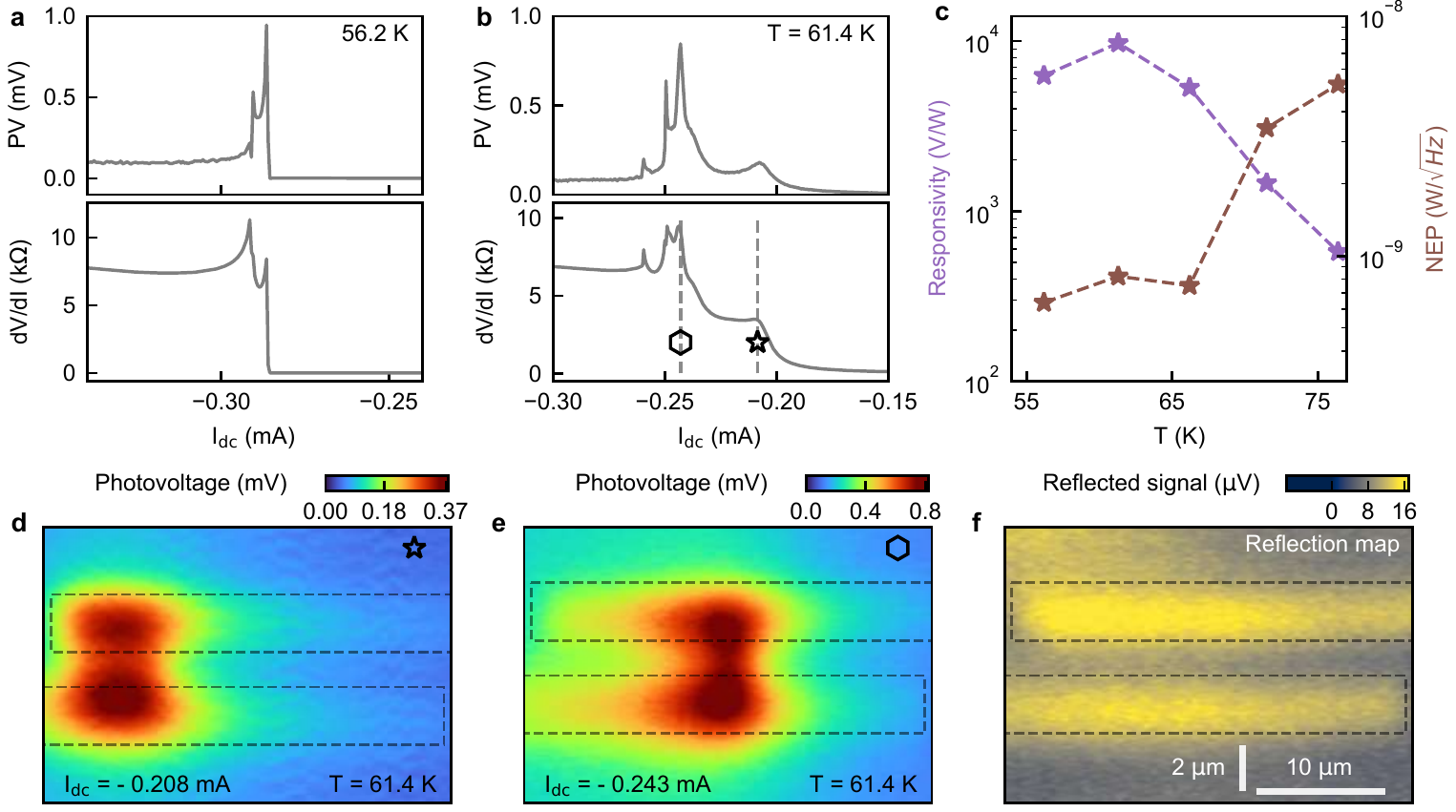}
\caption{ \label{fig:fig3} \textbf{Photoresponse of the long nanowire device (NW1) and photovoltage mapping along its length.} (a) $dV/dI$ (bottom panel) and simultaneously measured laser induced photovoltage across nanowire (top panel) as a function of \textit{dc} biasing current at 56.2 K. Laser was directly focused on the nanowire with a 20X objective during this measurement. \SI{1}{\micro\ampere} \textit{ac} excitation current of 170 Hz was added to the dc biasing current for $dV/dI$ measurement. Incident laser intensity was modulated at different frequency (37 Hz) and induced photovoltage was recorded at the same laser intensity modulation frequency. The photovoltage is maximum at the biasing current where the nanowire transits from superconducting to normal state. Similar measurements were performed at different temperatures. (b) $dV/dI$ measurement at 61.4~K shows peaks at multiple biasing currents which is also mimicked by generated photovoltage. This was recorded with 50X objective. (c) Temperature variation of responsivity, calculated from recorded photovoltage with 20X objective. Right axis shows noise equivalent power (NEP) of the nanowire. (d) and (e) show spatial photovoltage maps across the nanowire with 50X objective when it is biased at $-0.208$~mA, and $-0.243$~mA (marked by dashed lines in (b)), respectively. Position of the maximum photovoltage changes with biasing currents. (f) Simultaneously mapped reflection image of the nanowire. The dashed rectangles indicate the boundary of deposited Cr/Au lines. }

\end{figure*}

%------------------------------------------------------------------

In high-$T_\text{c}$ superconducting films there are several possible reasons for resistive jumps in $I-V$ or $I-dV/dI$ characteristics -- flux flow instability \citep{xiao_voltage_1998,xiao_coexistence_1998}, phase slip process \citep{agafonov_observation_2001,sivakov_josephson_2003,delacour_quantum_2012}, hotspot effects \citep{xiao_coexistence_1998,niratisairak_observation_2008}, fluctuating charge stripe domains \cite{bonetti_electronic_2004}, hotspot assisted suppression of edge barrier by the transport current \citep{lyatti_experimental_2018}, thermal runaway process\citep{maza_transition_2008}. 
In wide and thick nanowires, all of the above mechanisms can co-exist within the same current range. Resistive jumps in $dV/dI$ in our nanowire are predominantly caused by hotspots. We ruled out the other possibilities as they do not account for the experimental signatures seen in our nanowire. Flux flow instability can not explain the resistive steps as the observed resistivity jumps are very large compared to resistivity jumps predicted from flux flow \citep{xiao_voltage_1998}. In the case of fluctuating charge stripe domains, an associated telegraphic fluctuation \citep{bonetti_electronic_2004} appears in $R$ vs. $T$ which we do not see in our nanowire. Phase slip processes like, phase slip centers (PSC) can give rise to multiple resistive jumps. However, those arise only in 1D systems (W $\ll \xi$, where $\xi$ is coherence length). Our present scenario W $\gg \xi$, allows us to discard this possibility. However, 2D analog of PSC can appear in the wider superconducting film (W $\gg \xi$) as phase slip lines (PSL) \citep{sivakov_josephson_2003}. But experimental signatures of these phase slip processes \citep{delacour_quantum_2012}-- equal resistive jumps, and single excess current are missing in our nanowire. Moreover, the temperature evolution of $I_\text{s}$ is well described by the hotspot model (details in Supporting Information, Figure~S4). We thus conclude that the resistive jumps in our long nanowire are due to hotspots formation. We discuss later how these hotspots are spatially distributed along the long nanowire.

%-------------------------------------------------------------------

%-------------------------------------------------------------------

We now turn to the central part of the study, the photoresponse of the nanowire. To measure photoresponse, we biased the nanowire with \textit{dc} current. A 630~nm laser is focused onto the nanowire with a 20X objective. To exploit lock-in detection, we modulate the intensity of the laser at a frequency \textit{f} and measure the generated photovoltage (PV) across the nanowire at the same modulation frequency. We sweep the \textit{dc} biasing current (in both directions, shown in Figure~S6, Supporting Information) and simultaneously measure $dV/dI$ and the generated photovoltage across the nanowire. Figure~\ref{fig:fig3}a shows one such measurement at 56.2~K. The maximum PV occurs at a biasing current where the nanowire has a peak in $dV/dI$. The exposed area of the nanowire absorbs energy from the incident laser and locally increases the temperature. Having positive $dR/dT$ of the nanowire and constant current biasing across it, causes positive electrothermal feedback on the nanowire because of Joule heating. The generated heat eventually forces the superconducting nanowire to latch to a resistive state producing PV. Since the nanowire is most susceptible to any change in resistance near S-N transition, it has maximum PV at the transition. We measure PV at different temperatures and calculate responsivity defined as the generated PV normalized by the incident laser power. 

Figure~\ref{fig:fig3}c shows the temperature dependence of the responsivity along with the noise equivalent power (NEP). As the temperature decreases the responsivity of the device increases up to 62~K, beyond which it shows a downward trend. NEP, on the other hand, shows the opposite behavior to that of the responsivity. The nanowire has maximum responsivity of 7.59 $\times$ 10$^{3}$ V/W at 62~K (This temperature is close to the knee of the S-N transition; resistance of the nanowire drops to zero at 67 K with zero bias), which is close to the other high-$T_\text{c}$ bolometers \citep{berkowitz_low_1996,li_high_1993,kakehi_infrared_1998,mechin_suspended_1997,johnson_ybasub_1993,barth_epitaxial_1995,seifert_high-t_2021}. A comparison with the existing bolometers is drawn in the Supporting Information, Figure~S7. The responsivity of the detector can be improved further as discussed later. Here we note that in our calculation of the responsivity, we assume complete absorption of the light falling on the nanowire. However, in practice, the absorption is smaller, meaning the responsivity can be even higher than our calculation. If we correct for the absorption by the nanowire ($\sim$~78 \%, estimated from the reflection map of Figure~\ref{fig:fig3}f. This value is close to the earlier reported value \citep{memon_infrared_1994}), the responsivity can be as high as 9.7 $\times$ 10$^{3}$ V/W. Since the detector works on the bolometric principle, it can operate for a wider range of wavelengths (See Supporting Information, Figure~S7).

%-------------------------------------------------------------------

We note the NEP of our bolometer is comparatively higher than the earlier reported high-$T_\text{c}$ bolometers\citep{berkowitz_low_1996,li_high_1993,kakehi_infrared_1998,mechin_suspended_1997,johnson_ybasub_1993,barth_epitaxial_1995,seifert_high-t_2021}. The major sources of noise that can results in higher NEP, are phonon noise, Johnson noise, amplifier noise, and excess 1/f noise generated by the superconducting material \cite{verghese_feasibility_1992}. In our case, the theoretical lower bound ($\sim 10^{-12}$ W/$\sqrt{Hz}$) is currently limited by the phonon noise ($\sqrt{4k_\text{B}GT^2}$, where $G$ is the thermal conductance between the bolometer and the substrate). The NEP can be improved either by increasing responsivity or by reducing noise coming from different sources. The higher phonon noise in our bolometer is due to the ionic glass that we use as a substrate. A proper choice of substrate with lower $G$ value will increase the responsivity and also lower the phonon noise. The NEP that we experimentally observe in our device is large possibly due to other circuit noises.

%-------------------------------------------------------------------

Next, we turn to the spatial maps of the generated photovoltage along the nanowire. The bottom panel of Figure~\ref{fig:fig3}b shows $dV/dI$ as a function of biasing current at 61.4~K. We see multiple peaks in the $dV/dI$ characteristics, as mentioned earlier. The top panel shows PV generated across the nanowire with a laser, focused on the nanowire with a 50X objective. There are multiple associated peaks in PV for each peak in $dV/dI$. Based on this response, we bias the nanowire with a constant current near one of the maxima of photovoltage and do a photovoltage mapping along the length of the nanowire. The mapping is done by rastering the incident laser beam along the nanowire with the help of a X-Y piezo scanner on which the sample is mounted. We simultaneously collect the light reflecting off the nanowire. This provides a reflection image of the nanowire and ensures one-to-one mapping between PV and position on the nanowire. We used 50X objective in this case for better resolution in imaging.

Figure~\ref{fig:fig3}d shows a PV map along the nanowire at a biasing current of $-0.208$~mA, indicated by a dashed line in Figure~\ref{fig:fig3}b. To get this spatial map, we send a constant $dc$ current of $-0.208$~mA through the nanowire and a laser is focused onto it through a 50X objective. The generated photovoltage across the nanowire is measured with a lock-in amplifier at the modulation frequency of the laser intensity. The measurement is done while rastering the sample stage with the help of piezo scanners as discussed above. For details, see Supporting Information, Figure S11. Remarkably, we do not see uniform PV along the nanowire length. Instead, we see the maximum PV appears near the edge of the nanowire. Figure~\ref{fig:fig3}e shows the PV map of the same nanowire at a different biasing current (at $-0.243$~mA, indicated by a dashed line in Figure~\ref{fig:fig3}b). Figure~\ref{fig:fig3}f shows simultaneously recorded reflection image of the nanowire. The dashed boxes in all the images indicate the boundary of the two Cr/Au lines which define the nanowire. At $-0.243$~mA bias current also, we see maximum PV localized at a region on the nanowire. However, in contrast to $-0.208$~mA, the PV maxima at $-0.243$~mA is near the middle of the nanowire length. This shows that the location of the maximum photovoltage along the nanowire length depends on the biasing current. Photovoltage maps at other temperatures are shown in Supporting Information (Figure~S8). 

Here, we note two interesting points. First, the photoresponse maps reveal a substantial amount of photovoltage arising across the nanowire even when the laser is focused onto the Cr/Au lines (marked by dashed rectangles), away from the nanowire. This suggests that the light absorbed by the Cr/Au covered region is converted to heat which is then laterally transferred to the nanowire. The responsivity of the detector can be further improved by carefully engineering its design, combining light-absorbing materials as a heat transducer to the nanowire. Second, we map photovoltage along the nanowire at different temperatures (shown in Supporting Information, Figure~S8) and notice the spot of maximum photovoltage also changes its spatial position along the nanowire with change in temperature. This suggests there can be other possible mechanisms at play that we discussed earlier apart from the hotspots formation, which can account for the dynamics of these hotspots along the nanowire.  

%--------------------------------------------------------------------

\begin{figure}
\includegraphics[width=7.6cm]{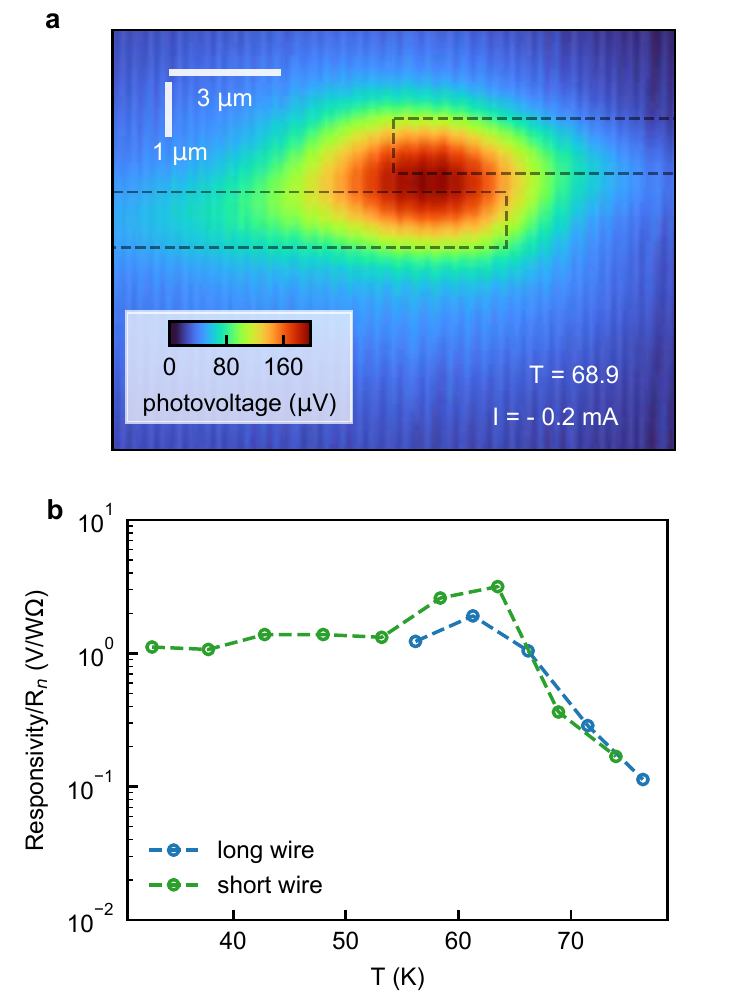}
\caption{ \label{fig:fig4}  \textbf{Photoresponse of short nanowire (NW2).} (a) Photovoltage map of the nanowire at 68.9~K. During the mapping the nanowire was biased with $-0.2$~mA current. The dashed rectangles mark the boundary of the deposited Cr/Au lines. (b) Responsivity of both the long and short nanowires normalized by their normal state resistance $R_\text{n}$ over a large temperature range.}
\end{figure}

%--------------------------------------------------------------------

We now contrast our results of photoresponse for the short nanowire (NW2), made out of 4 unit cells (12~nm) thick BSCCO, with that of the long nanowire. Unlike the long nanowire (NW1), photovoltage mapping of the short nanowire (NW2) shows only one spot of photovoltage maxima (uniform along the length), as shown in Figure~\ref{fig:fig4}a. The dashed rectangles in Figure~\ref{fig:fig4}a are the boundary of the deposited Cr/Au lines. The long nanowire showed localized spots of maximum photovoltage associated with each peak in $dV/dI$ at different locations along its length. For the long nanowire, this suggests presence of weak spots (with different switching currents) along the length due to nonuniform width distribution. The nonuniformity can happen because of inhomogenous Cr deposition over the BSCCO flake. The maximum photovoltage appears in a region along the long nanowire at a biasing current that corresponds to the switching current of that region. The inhomogeneity is less probable for the shorter version of the nanowire, and we achieve nanowire of uniform width. For more discussion on nanowire length-dependent inhomogeneity of Cr deposition, see Supporting Information (Figure~S2). We also measured the photoresponse of the short nanowire (NW2) at different temperatures following the previously described method. We compare the responsivity of the long nanowire (NW1) with that of the short nanowire (NW2) by normalizing with their respective normal state resistance $R_\text{n}$ in Figure~\ref{fig:fig4}b. It shows the responsivity to be proportional to the normal state resistance of the nanowire.  

%------------------------------------------------------------------

%-------------------------------------------------------------------

\begin{figure*}
\includegraphics[width=15.5cm]{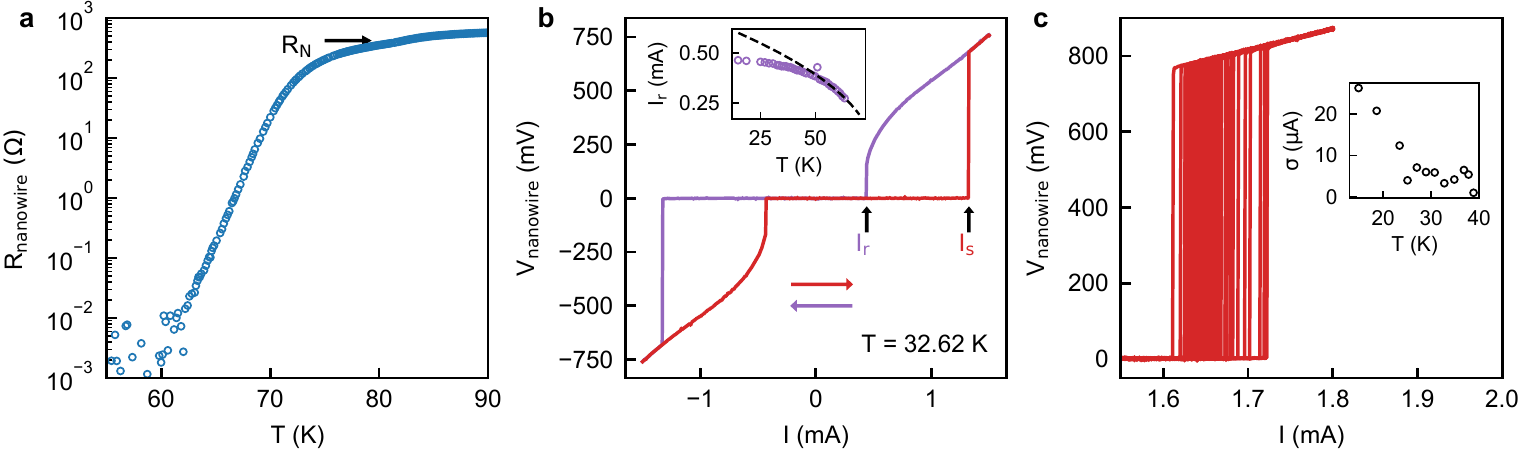}
\caption{ \label{fig:fig5}  \textbf{Electrical characterization of short superconducting nanowire.} (a) Four-probe resistance versus temperature of the nanowire in logarithmic scale. Normal state resistance of the nanowire is indicated by the arrow where entire BSCCO flake has gone superconducting except the nanowire. Thickness of the BSCCO flake used to fabricate this nanowire was 12~nm (4 unit cells) thick. (b) \textit{DC} $I-V$ characteristic of the nanowire at 32.62~K. The two different coloured line plots are for two different sweep directions of the biasing current indicated by the arrows. $I_\text{s}$ and $I_\text{r}$ are switching and retrapping currents, respectively. Inset shows  evolution of $I_\text{r}$ with temperature. The black dashed line is the fit of $I_\text{r}$ with hotspot model. The experimental data points significantly deviates from hotspot model and satuarates at low temperatures. (c) Switching events of the nanowire at 18.6~K for multiple sweeps. The stochastic nature of the $I_\text{s}$ is clearly seen as switching from superconducting to normal state at different currents. Inset shows standard deviation of the $I_\text{s}$ distribution with temperature.}

\end{figure*}

Next, we carefully study the transport of the short nanowire (NW2) with smaller cross-section ($\sim~3600~nm^{2}$) at high biasing currents. Figure~\ref{fig:fig5}a shows the $R$ vs. $T$ of the NW2. The normal state resistance $R_\text{n}$ of the nanowire (420~$\Omega$) is indicated by the arrow where the part of the BSCCO except the nanowire has become superconducting. 

%-------------------------------------------------------------------

%------------------------------------------------------------------

Figure~\ref{fig:fig5}b shows $I-V$ characteristics of the short nanowire (NW2) at 32.62~K for biasing current swept in both directions. In contrast to NW1, we do not see multiple steps in $I-V$ characteristics which again emphasizes the uniformity of the nanowire. However, there is pronounced hysteresis with the sweep direction. $I-V$ characteristics at different temperatures are shown in the Supporting Information (Figure~S9). We see the nanowire has a bistable state between $I_\text{s}$ and $I_\text{r}$ depending on the history of sweep direction of the biasing current. To understand the hysteresis, we plot $I_\text{r}$ of the nanowire as a function of temperature in the inset of Figure~\ref{fig:fig5}b. The black dashed line overlaid on the inset figure is $I_\text{r}$, as predicted by the hotspot model. Although experimental values of $I_\text{r}$ follow the model quite well at higher temperatures, the values significantly deviate at low temperatures and tend to saturate. The fact that $I_\text{r}$ is almost temperature independent at low temperatures suggests that at low temperatures, hysteresis is not explained by the hotspot model and that there can be other mechanisms. This is the first signature that could point to phase slip events in NW2 \citep{lyatti_experimental_2018,michotte_condition_2004}. These can be thermal \citep{langer_intrinsic_1967,mccumber_time_1970,agafonov_observation_2001} or quantum \citep{giordano_evidence_1988,
 bezryadin_quantum_2000,lau_quantum_2001} phase slips that occur along a uniform nanowire where the superconducting order parameter momentarily vanishes (accompanied by a 2$\pi$ phase slip across it) because of fluctuations, giving rise to non-zero resistance below superconducting transition temperature $T_\text{c}$.

%-------------------------------------------------------------------

%-------------------------------------------------------------------

Additionally, we find that $I_\text{s}$ is stochastic in nature, while we do not see the same behavior for $I_\text{r}$. This means that at a fixed temperature, the switching current varies from one sweep to another resulting in a distribution of the switching current. This stochasticity of switching was first studied by Fulton and Dunkleberger in Josephson junctions \citep{fulton_lifetime_1974}. It has been later shown that statistics of $I_\text{s}$ can provide important information about the fluctuation phenomenon present in the systems \citep{sahu_individual_2009}. To study this effect, we take 100 switching events at different temperatures. The result at 18.6~K is shown in Figure~\ref{fig:fig5}c, where we see different $I_\text{s}$ at different runs. Switching distributions at different temperatures are shown in Supporting Information (Figure~S10). One would expect the distribution width of $I_\text{s}$ to scale with the thermal noise, which means the width should decrease as the temperature is reduced \citep{fulton_lifetime_1974}; at low temperature, it should saturate where thermal fluctuations are frozen out, and only quantum fluctuations are left \citep{martinis_experimental_1987}. To see how this switching distribution evolves with temperature, we plot the standard deviation $\sigma$ of the distribution with temperature in the inset of Figure~\ref{fig:fig5}c. We see the switching distribution width increases instead as the temperature is reduced. This behavior has been seen previously in nanowires \citep{sahu_individual_2009} and indicates two possibilities -- an increase of distribution width at low temperatures by multiple thermal phase slips or the presence of quantum phase slips. A more important quantity to look for in this aspect is the rate of phase slip processes. From the experimentally obtained switching current distribution we transform the data to the switching rate. Next, we calculate the thermally activated phase slip (TAPS) rate ($\Gamma_{TAPS}$) using the overheating model that allows only thermally activated processes, and it agrees well with the experiment (shown in Supporting Information, Figure~S10). Taken together, the two evidences suggest the presence of multiple thermal phase slips in short nanowire (NW2) with a smaller cross-section at low temperatures.

%-------------------------------------------------------------------

%------------------------------------------------------------------
\section{Conclusion}

In conclusion, we made nanowires of high-$T_\text{c}$ superconductor with different aspect ratios. The long nanowire (700~nm wide) shows bias current-dependent hotspots of photovoltage along the length. The short nanowire with a smaller cross-section indicates the presence of multiple thermally activated phase slips at low temperatures. We show that as-fabricated nanowires can be used as a sensitive detector of photons in the visible wavelength. The responsivity of the detector scales with the normal state resistance of the nanowire suggesting that the responsivity can be improved by increasing the length, reducing the width of the nanowire, and reducing the thickness of the superconducting flake. Based on this architecture, a heterostructure can be realized by transferring efficient light absorbing van der Waals materials \cite{jariwala_near-unity_2016,epstein_near-unity_2020} onto the nanowire as a heat transducer to further increase the responsivity of the detector. The method can be employed to make superconducting nanowires of even smaller cross-section to host quantum phase slips that could allow realization of superconducting nanowire based qubits \citep{mooij_phase-slip_2005,mooij_superconducting_2006} and single photon detectors \citep{merino_two-dimensional_2023,charaev_single-photon_2022}.

%--------------------------------------------------------------------

\section*{References}

\bibliography{Nanowire_bolometer_2}

\section*{Acknowledgments}
We thank A. Thamizhavel and Ruta N. Kulkarni for their help in growth of the BSCCO single crystals. We thank Pratap Chandra Adak, Supriya Mandal, and Subhajit Sinha for their comments on the manuscript. We thank Frank Zhao for pointing out the idea of using vacuum grease to attach SiN mask with substrate for better resolution of the deposited pattern by shadow mask evaporation. We also thank Amit P. Shah for help in making SiN membrane, Bhagyashree A. Chalke, Rudheer D. Bapat, and Jayesh B. Parmar for doing SEM imaging of the devices. We acknowledge Department of Science and Technology (DST), Nanomission grant SR/NM/NS-45/2016, CORE grant CRG/2020/003836, and Department of Atomic Energy (DAE) of Government of India for support.

\section*{Author contributions}
S.G. fabricated the devices, did  measurements and analyzed the data. D.A.J. grew the BSCCO crystals. S.G. and M.M.D. wrote the manuscript. All authors provided inputs to the manuscript. M.M.D. supervised the project.

\section*{Competing interests}
The authors declare no competing financial interests.

\clearpage

%........................Supplementary Information..................%

\widetext
\begin{center}
\textbf{\large Supporting Information: Nanowire bolometer using a 2D high-temperature superconductor}
\end{center}

\setcounter{equation}{0}
\setcounter{figure}{0}
\setcounter{table}{0}
\setcounter{section}{0}
\setcounter{page}{1}

\renewcommand{\thesection}{S\arabic{section}}
\renewcommand{\theequation}{S\arabic{equation}}
\renewcommand{\thefigure}{S\arabic{figure}}
\renewcommand{\thepage}{S\arabic{page}}

 \makeatletter
\def\@fnsymbol#1{\ensuremath{\ifcase#1\or \dagger\or *\or \ddagger\or
   \mathsection\or \mathparagraph\or \|\or **\or \dagger\dagger
   \or \ddagger\ddagger \else\@ctrerr\fi}}
    \makeatother

\section{Optical and SEM micrograph of nanowires}

\begin{figure}[h]
\includegraphics[width=15.5cm]{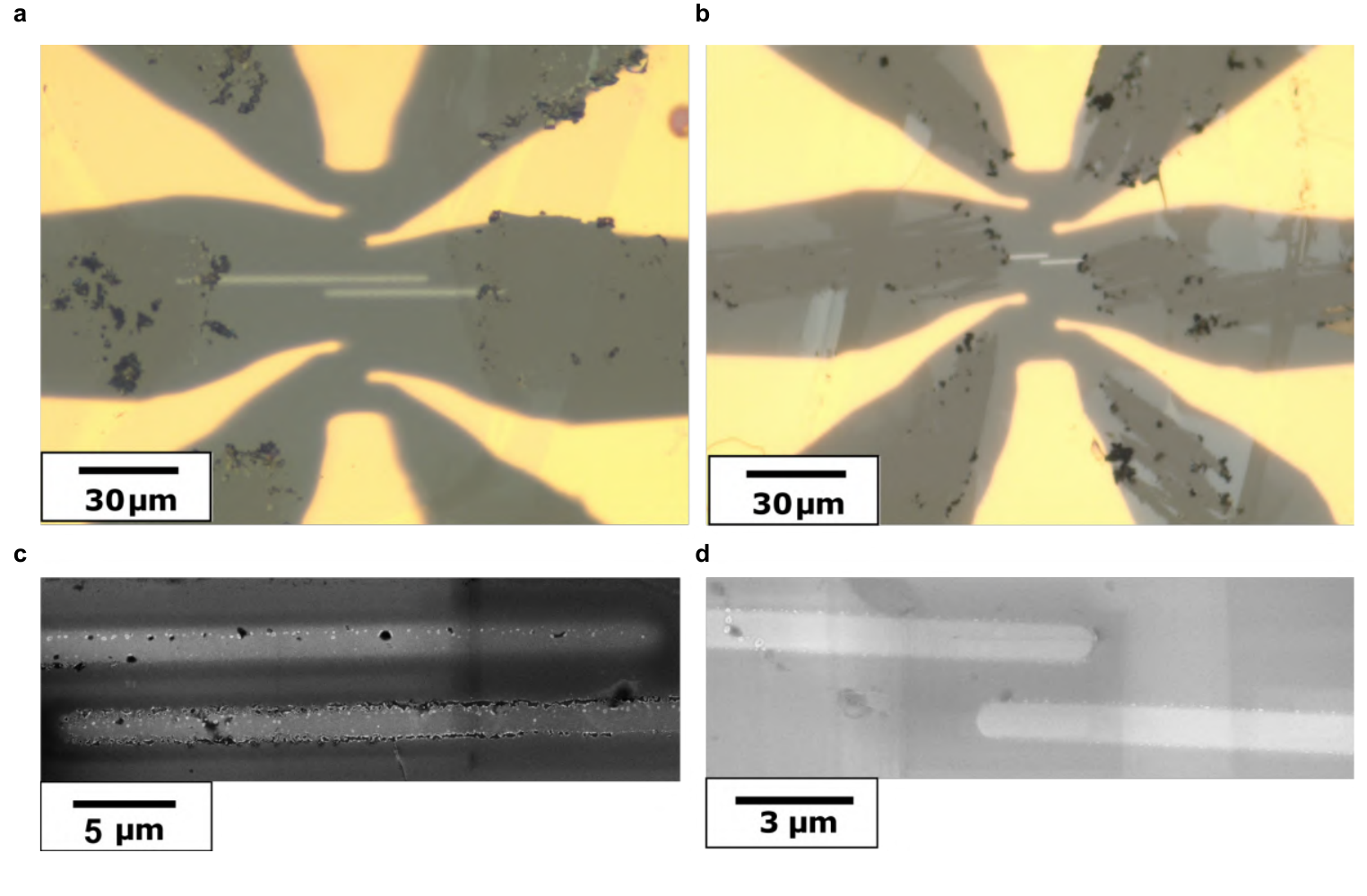}
\caption{ \label{fig:figS1}{\footnotesize Optical and SEM micrograph of nanowires. (a) Optical image of long nanowire (NW1) with length \SI{30}{\micro\metre}, width 700~nm and thickness 9~nm. The unwanted part of BSCCO flake was removed as discussed in \citep{ghosh_-demand_2020}. (b) Optical image of short nanowire (NW2) device with length \SI{3}{\micro\metre}, width 300~nm and thickness 12~nm. (c) and (d) are SEM images of NW1 and NW2 respectively after all the measurements are done. }}
\end{figure}

\section{$R$ vs. $T$ of nanowires with different widths}

\begin{figure}[h]
\includegraphics[width=15.5cm]{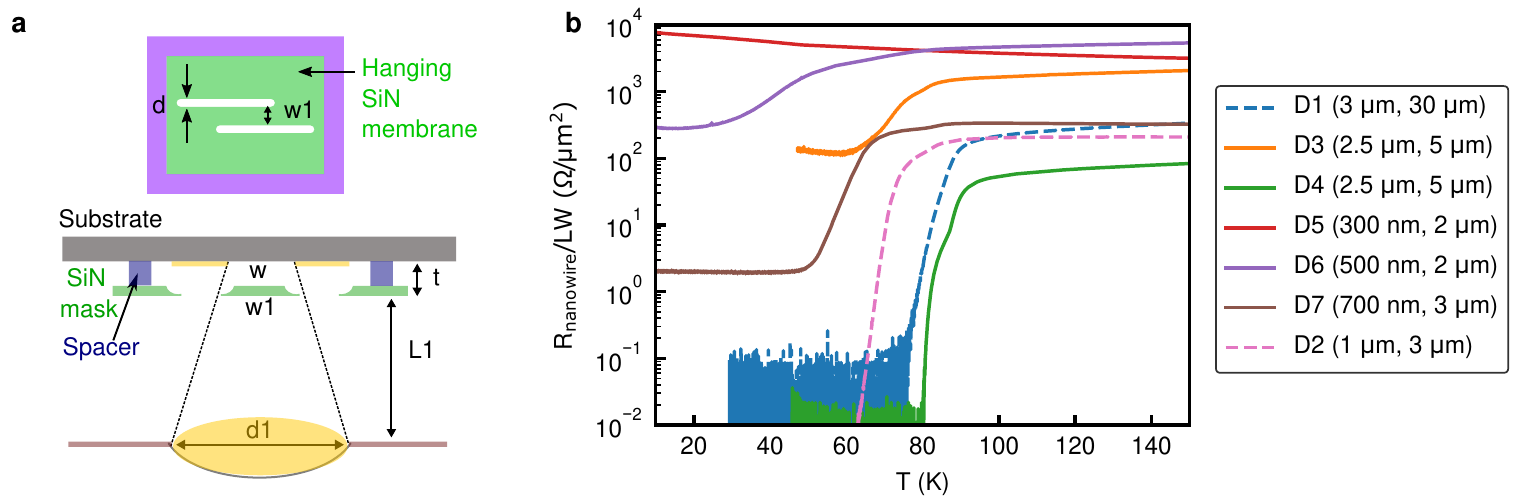}
\caption{ \label{fig:figS2} {\footnotesize Spreading of feature size during evaporation through mask and $R$ vs. $T$ of nanowires with different widths and lengths. (a) Schematic of SiN mask for nanowire geometry. Two slits with width $d$ are etched out from hanging SiN membrane with $w1$ separation between them. The separation $w1$ reduces to $w$ during Cr deposition ($d1$ is the size of the evaporating source metal) because of spreading which is determined by thickness of spacer. (b) $R$ vs. $T$ of nanowires with different widths and lengths. The widths and lengths are specified as two numbers in the parentheses in legends. The widths mentioned here (first number in the parentheses of legends) are the separation $w1$ in the mask pattern. The actual widths of the resulting nanowires are smaller. The reduction in widths are determined by the thickness of spacer. The dashed curves are for long (NW1) and short (NW2) nanowires presented in the main text. }}
\end{figure}

Figure~\ref{fig:figS2}a top panel shows schematic mask pattern for nanowire structure on a hanging SiN membrane (thickness 300 nm) where two slits of width $d$ separated from each other by $w1$ are etched out by reactive ion etching. The resulting mask is then aligned with the BSCCO flake. Bottom panel of Figure~\ref{fig:figS2}a shows schematic for metal deposition through the SiN mask aligned and attached on the substrate having flakes of BSCCO exfoliated on it. For attaching the SiN mask on the substrate we used two different methods. In one case we used a double sided adhesive Kapton tape which acts as the spacer. If we take the separation between mask and substrate to be $t$, the resulting gap between the two Cr lines deposited through the mask can be approximated as $w \approx w1 - (d1/L1)t$. Therefore, the spreading of feature size or reduction in separation between the two Cr lines after evaporation is proportional to $t$. To reduce the spreading, we used vacuum grease for attaching the mask with the substrate in the second method.

Figure~\ref{fig:figS2}b shows $R$ vs. $T$ of nanowires with different widths and lengths. For devices D1, D3 and D4, double sided adhesive Kapton tape was used to attach mask on the substrate. For devices D5, D6, D7 and D2 vacuum grease was used to attach mask on the substrate. Because of finite thickness of the Kapton tape, the spreading was more in the first set of devices (This can also lead to inhomogeneity of Cr deposition). As seen from the plots, depending on the separation between the two Cr lines, the behaviour of the nanowires evolves from being superconducting to insulating. The actual widths of the nanowires are estimated as an upper bound from their $R$ vs. $T$ behaviour. As an example, when using vacuum grease as the spacer, upto $\sim~700$ nm gap (D7) between two Cr lines (in the SiN mask) does not lead to a nanowire that becomes fully superconducting. Whereas with a gap of \SI{1}{\micro\metre}, the resulting nanowire becomes superconducting which gives an upper limit of $\sim~300$ nm     on the actual width of the nanowire.

\section{Evolution of dV/dI with temperature of nanowire and pristine BSCCO }

\begin{figure}[h]
\includegraphics[width=14cm]{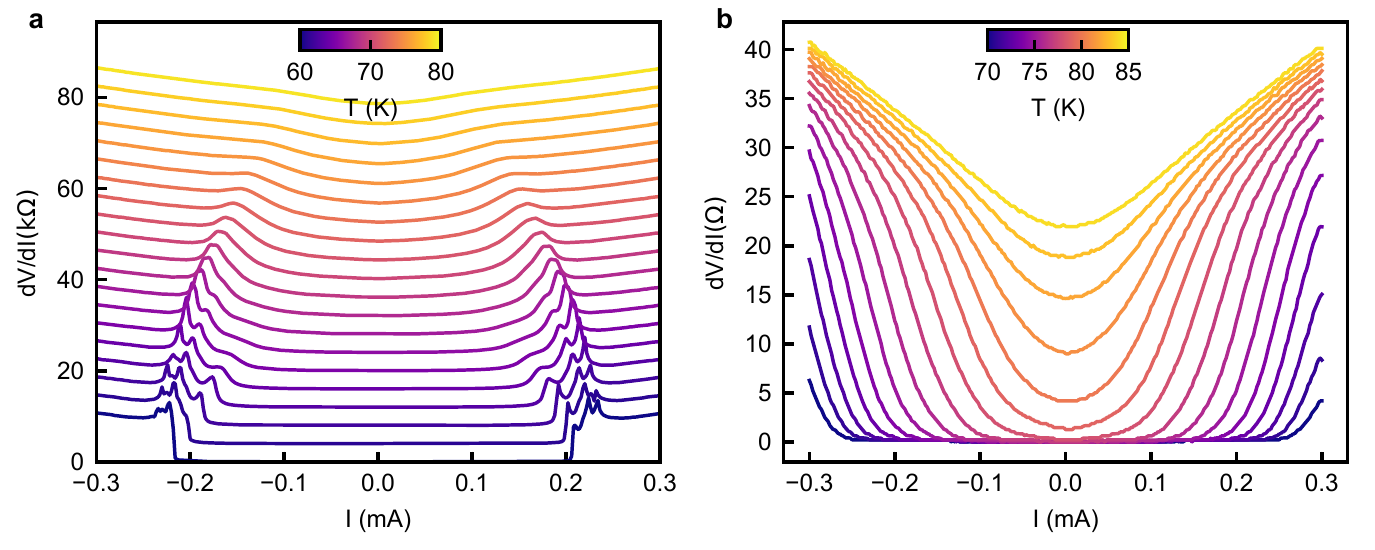}
\caption{ \label{fig:figS3} {\footnotesize $dV/dI$ as a function of biasing current of long nanowire (NW1) and pristine BSCCO at different temperatures. (a) $dV/dI$ of nanowire at different temperatures. Different coloured curves are for different temperatures as mapped by colour bar. For clarity the curve for each temperature is shifted in y axis by a constant amount. (b) $dV/dI$ across pristine BSCCO at different temperatures. Because of large critical current across pristine BSCCO (because of large cross section), the entire superconducting to normal transition curve is not produced at all temperature as we restricted large current to flow through the nanowire so that it does not get damaged.}}
\end{figure}

\section{Evolution of switching current with temperature}

\begin{figure}[h]
\includegraphics[width=14cm]{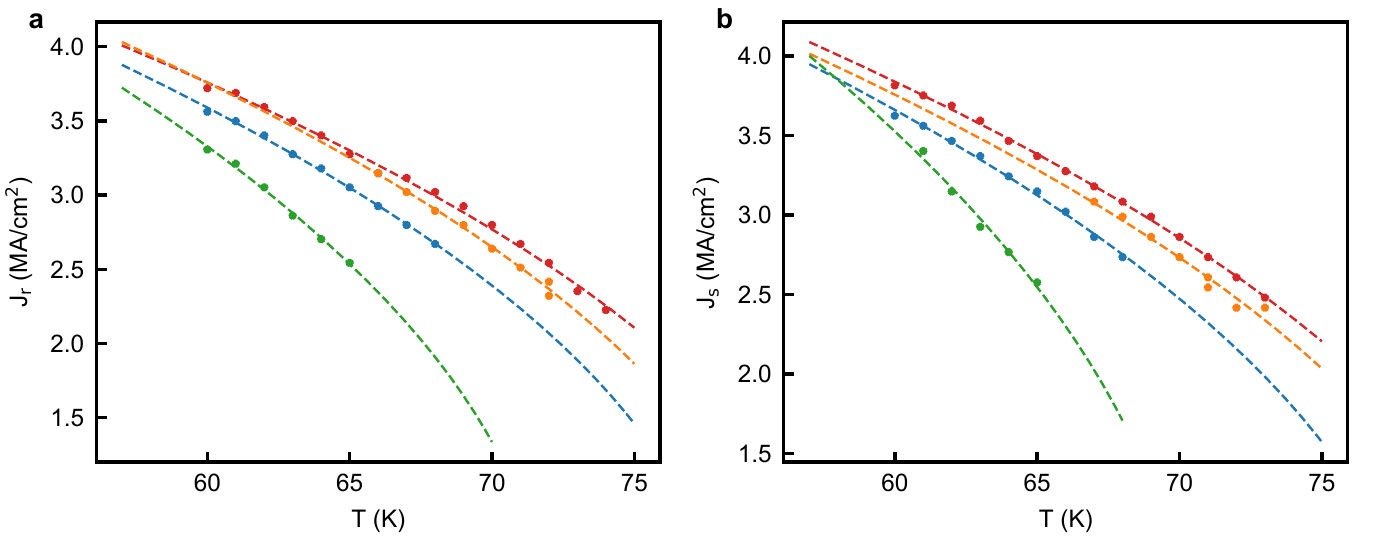}
\caption{ \label{fig:figS4}{\footnotesize Switching and retrapping current evolution with temperature of long nanowire (NW1). As seen from Figure~\ref{fig:figS3}a there are multiple transitions at each temperature for both switching and retrapping branch. (a) and (b) are evolution of those retrapping and switching current densities with temperatures and their fit (dashed lines) with hotspot model as discussed in the main text. }}
\end{figure}

\section{Shunting effect on hysteresis of nanowire }

\begin{figure}[h]
\includegraphics[width=15.5cm]{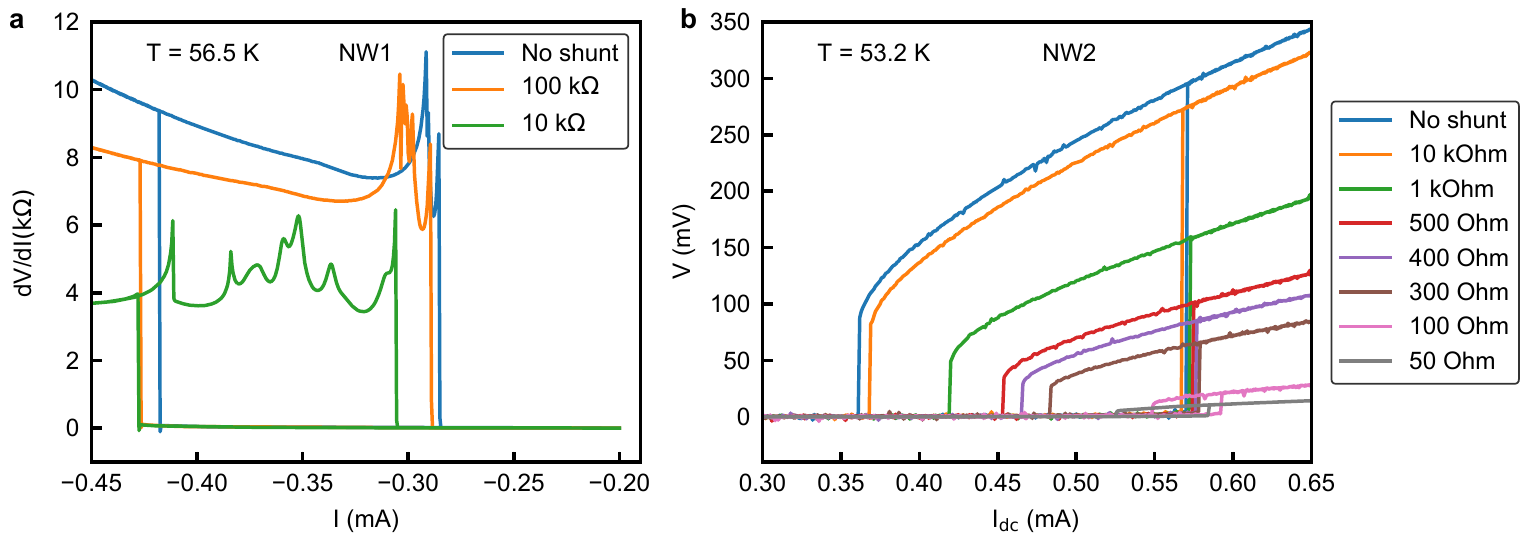}
\caption{ \label{fig:figS5} Shunting effect on reducing hysteresis in the nanowires. (a) $dV/dI$ vs. $I$ of long nanowire (NW1) at 56.5~K with different shunting resistances. With decreasing shunting resistance hysteresis between up and down sweep of biasing current reduces and number of peaks in $dV/dI$ in the retrapping branch increases. (b) $I-V$ of short nanowire (NW2) at 53.2~K with different shunting resistances.  }
\end{figure}

\section{Photovoltage in both direction of biasing current sweep}

\begin{figure}[h]
\includegraphics[width=14cm]{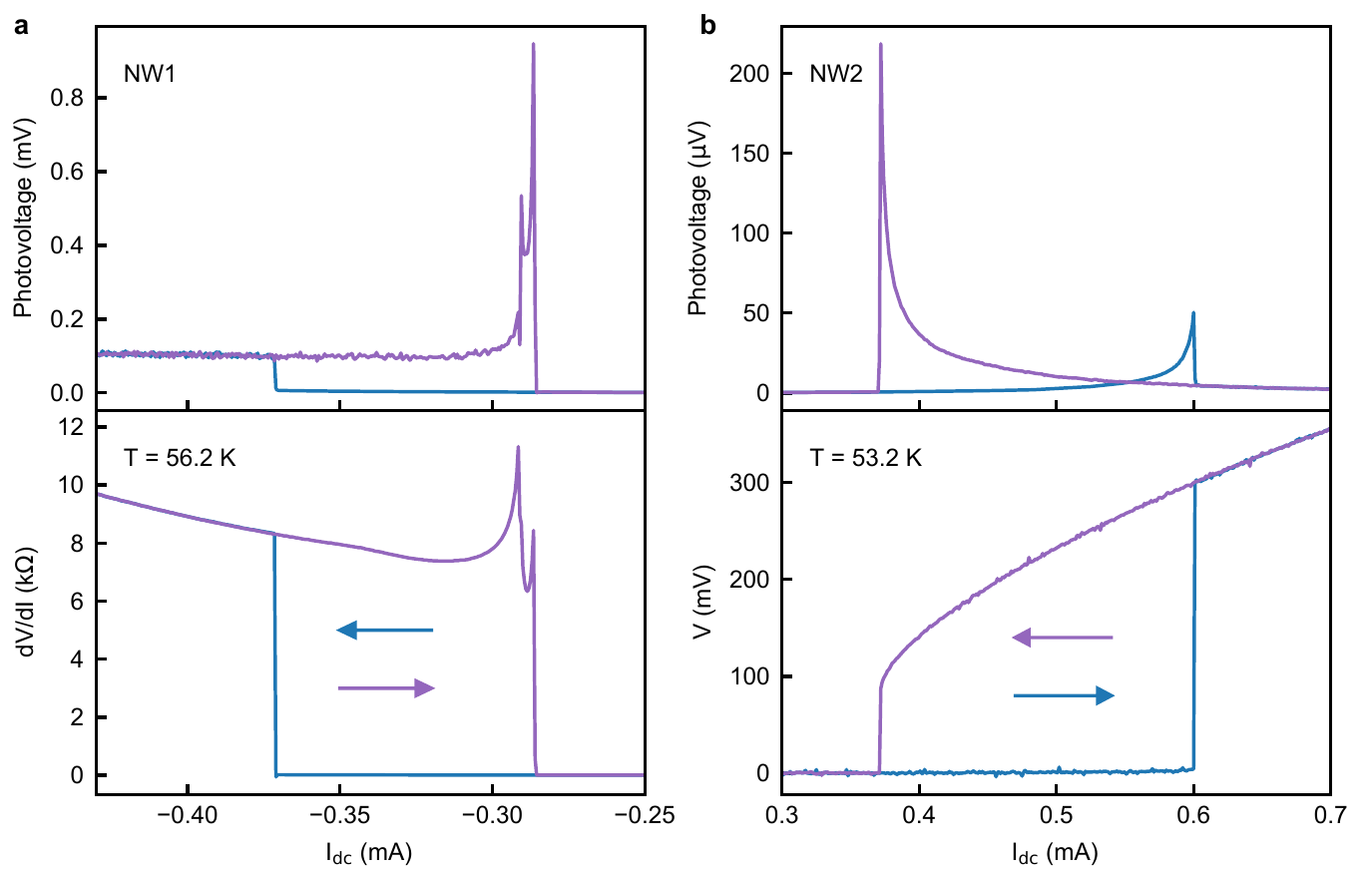}
\caption{ \label{fig:figS6}{\footnotesize Photoresponse in switching and retrapping branch of superconducting to normal transition. (a) $dV/dI$ and simultaneously measured photoresponse of long nanowire (NW1) at 56.2~K in both switching and retrapping branch. The arrows indicate sweep direction of biasing current. Photovoltage is maximum in the retrapping branch. (b) $I-V$ and simultaneously measured photovoltage of short nanowire (NW2) at 53.2~K.}}
\end{figure}

\section{Responsivity comparison with existing detectors}

\begin{figure}[h]
\includegraphics[width=14cm]{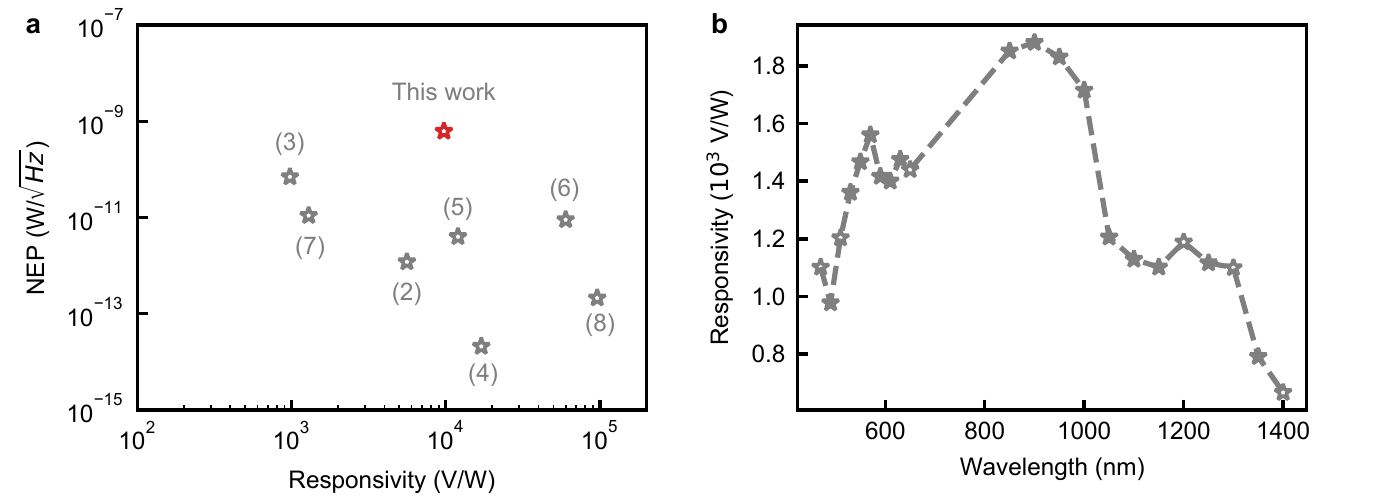}
\caption{ \label{fig:figS7}{\footnotesize (a) Comparison of our detector's performance with the existing high-$T_\text{c}$ superconducting infrared detectors \citep{berkowitz_low_1996,li_high_1993,kakehi_infrared_1998,mechin_suspended_1997,johnson_ybasub_1993,barth_epitaxial_1995,seifert_high-t_2021}. (b) Wavelength dependent responsivity of the detector. }}
\end{figure}

\section{Photovoltage mapping of long nanowire (NW1) at different temperatures}

\begin{figure}[h]
\includegraphics[width=12.7cm]{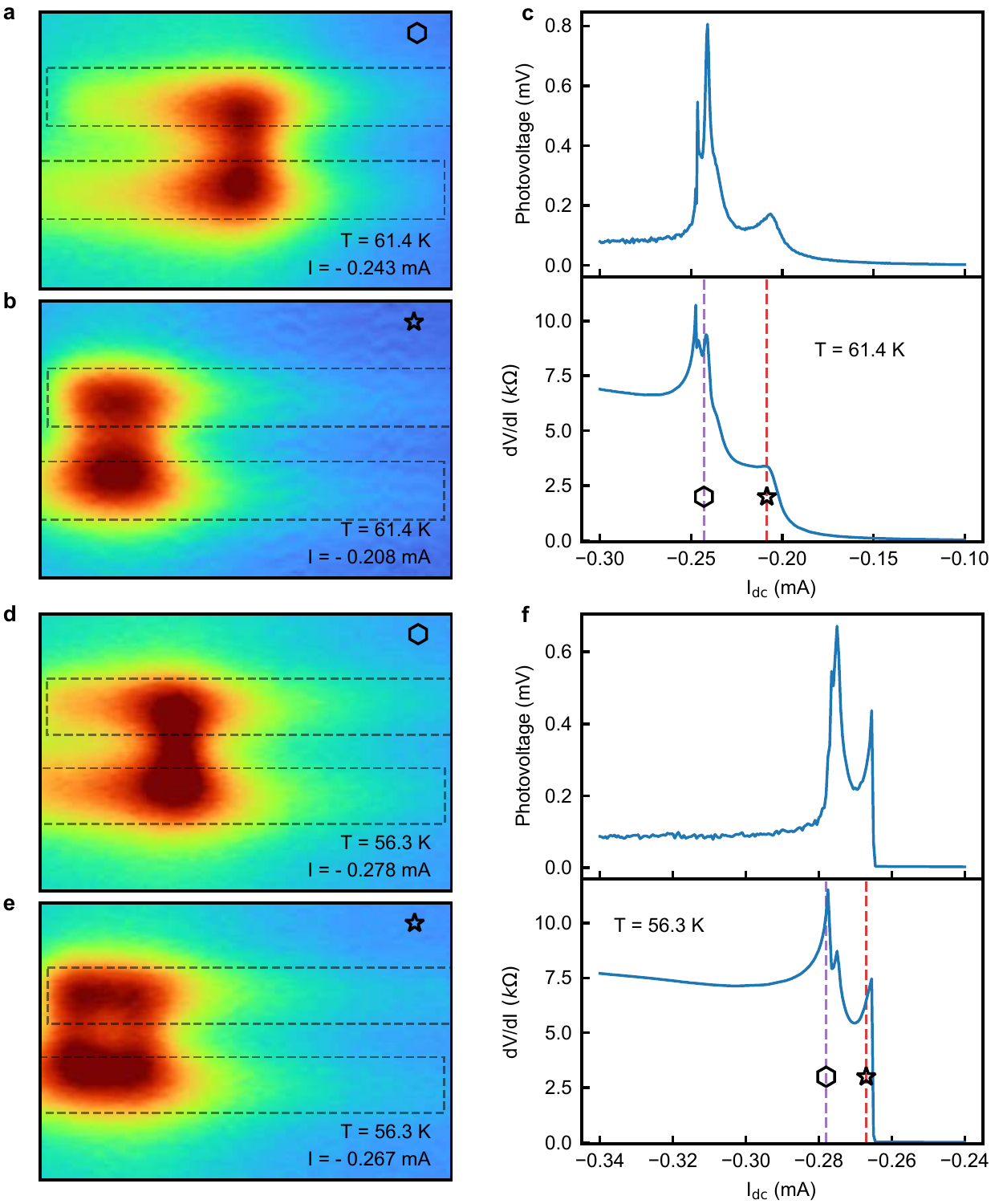}
\caption{ \label{fig:figS8}{\footnotesize Spatial mapping of photovoltage of long nanowire (NW1) at different temperatures. (a) and (b) Photovoltage mapping at 61.4~K (presented in the main text) while the nanowire is biased at -0.243~mA and -0.208~mA respectively as indicated by dashed lines in (c). (d) and (e) are photovoltage mapping of the same nanowire at 56.3~K while it is biased at -0.278~mA and -0.267~mA respectively, indicated by dashed lines in (f). As seen from the photovoltage mapping, the position of the photovoltage maxima along the nanowire is also function of temperature. }}
\end{figure}

\section{$I-V$ of short nanowire (NW2) at different temperatures}

\begin{figure}[h]
\includegraphics[width=14cm]{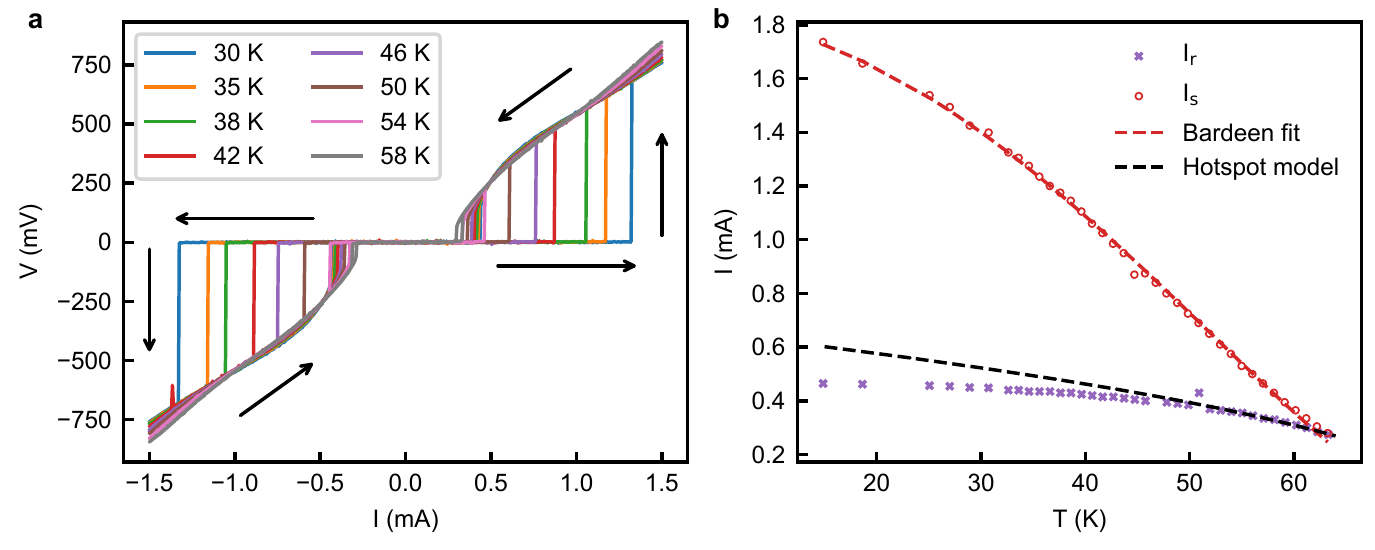}
\caption{ \label{fig:figS9}{\footnotesize (a) $I-V$ of short nanowire (NW2) at different temperatures. (b) Switching and retrapping current evolution with temperature. The switching current $I_\text{s}$ is not fitted with the hotspot model in contrast to the long nanowire (NW1). Rather it follows quite well with the Bardeen theory \citep{bardeen_critical_1962} . $I_\text{r}$ on the other hand follows hotspot model at high temperatures and deviates from it at low temperatures as mentioned in the main text as well. }}
\end{figure}

\section{Switching statistics and rate at different temperatures of short nanowire (NW2)}

At a fixed temperature $T$ and at a fixed bias current $I$ the thermally activated phase slip (TAPS) rate is given by,

\begin{equation}\label{1}
\Gamma_{TAPS}(T,I)= \frac{\Omega_{TAPS}(T)}{2\pi} \exp(-\frac{\Delta F(T,I)}{k_{B}T})
\end{equation}

Where $\Omega_{TAPS}(T)$ is attempt frequency, $\Delta F(T,I)$ is the free energy barrier for phase slips at bias current $I$ and $k_B$ is the Boltzmann constant.

The attempt frequency is given by,

\begin{equation}
\Omega_{TAPS}(T) = \frac{L}{\xi(T)}\sqrt{\frac{\Delta F(T)}{k_{B}T}}\frac{1}{\tau_{GL}}
\end{equation}

Here, $L$ is the length of the nanowire, $\xi(T) = \xi(0)/\sqrt{(1-T/T_c)}$ is the coherence length, $\tau_{GL}$ is the Ginzburg-Landau (GL) relaxation time, and $\Delta F(T) = (8\sqrt{2}/3)(H_{c}^2(T)/8\pi)A\xi(T)$ is free energy barrier for phase slip. $H_c(T)$ and $A$ are respectively the critical field and cross-sectional area of the nanowire.

The free energy barrier $\Delta F(T)$ can be alternatively expressed as \cite{tinkham_quantum_2002,tinkham_hysteretic_2003,langer_intrinsic_1967,bardeen_critical_1962},

\begin{equation}
\Delta F(T) = \frac{\sqrt{6} \hbar I_c(T)}{2e}
\end{equation}

Where $I_c(T)$ is the depairing current and is given by,

\begin{equation}
I_c(T) = (92 \mu A)\frac{LT_c}{R_n \xi (0)} \left(1-\left(\frac{T}{T_c}\right)^2\right)^{3/2}
\end{equation}

Here $L$ and $\xi(0)$ are in nm, $T_c$ is in K and normal state resistance of the nanowire $R_n$ is in Ohm.

Now the free energy barrier which takes into account both the temperature $T$ and high bias current $I$ can be written as \cite{tinkham_hysteretic_2003,langer_intrinsic_1967},
,

\begin{equation}
\Delta F(T,I) = \frac{\sqrt{6} \hbar I_c(T)}{2e} \left(1-\frac{I}{I_c}\right)^{5/4}
\end{equation}

Putting all the expressions together we arrive at the following expression of the $\Gamma_{TAPS}$,

\begin{equation}
\Gamma_{TAPS}(T,I)= \left(\frac{1}{2\pi}\right) \left(\frac{L}{\xi(T)}\right) \left(\frac{1}{\tau_{GL}}\right) \left(\frac{\Delta F(T)}{k_{B}T}\right)^{1/2} \exp(-\frac{\Delta F(T,I)}{k_{B}T})
\end{equation}

We use this above expression to estimate the phase slip rate caused by thermal fluctuation and compare with the experimentally obtained switching rate.

From experiment we obtain counts of different switching currents ($I_s$) at a fixed temperature. Figure~\ref{fig:figS10}a shows the counts of switching currents at different temperatures. From the count we transform the data to get probability density of switching current $P(I_s)$ by using the relation $count = P(I_s) \times \Delta I \times N $, where $N$ is the number of total measurements (100) and $\Delta I$ is the bin size of biasing current (1 $\mu A$) chosen for plotting the switching current distribution. Following Fulton-Dunkleberger analysis \citep{fulton_lifetime_1974} we then transform $P(I_s)$ into information on the switching rates ($\Gamma_{s} (T,I)$) at fixed current and temperature. To calculate the rate from the probability density of switching current, we split the current axis into bins of size $\Delta I$ and the corresponding current values are numbered $I_k = I_c - k\Delta I$, where the integer bin number $k$ obeys $0<k<N$, with the highest bin number $N$ defined via $N = I_c/\Delta I$. Here $I_c$ is the maximum value of the switching current for a particular distribution. The switching rate then can be calculated as \citep{bezryadin_superconducting_2010},

\begin{equation}
\Gamma_{k} = \Gamma (I_k) = \left(\frac{dI}{dt}\right) \left(\frac{1}{\Delta I}\right) ln\left( \frac{\sum_{i=0}^k P(I_i)}{\sum_{i=0}^{k-1} P(I_i)} \right)
\end{equation}

Here, $dI/dt$ is the rate of sweeping biasing current during the measurement. Figure~\ref{fig:figS10}c shows the calculated rate of switching at different temperatures (open circles).

Now to calculate the thermal phase slip rate we first fit the switching current with the expression of depairing current $I_c$ from (4) using the following known parameters of the nanowire : $L = 3~\mu m$, $R_n = 420~\Omega$, and $\xi(0) = 3.2~nm$ \citep{naughton_orientational_1988}. And as a fitting parameter we obtain the critical temperature $T_c = 73.2~K$. As we see from Figure~\ref{fig:figS10}b, the switching current $I_s$ is $\sim~8$ times smaller than the depairing current $I_c$. We rewrite (4) as $I_c(T) = I_0 (1-(T/T_c)^2)^{3/2}$ which is the Bardeen foumula \citep{bardeen_critical_1962}. Next we use this $I_0$ which fits the experimentally obtained $I_s$ and plug the values into (6) to get $\Gamma_{TAPS}(T,I)$. For $1/\tau_{GL}$, we take a typical value of $\sim 10^{12}$. $\Gamma_{TAPS}(T,I)$ are then calculated and plotted as solid lines in Figure~\ref{fig:figS10}c which matches quite well with experiment at all temperatures.

\begin{figure}[h]
\includegraphics[width=12.7cm]{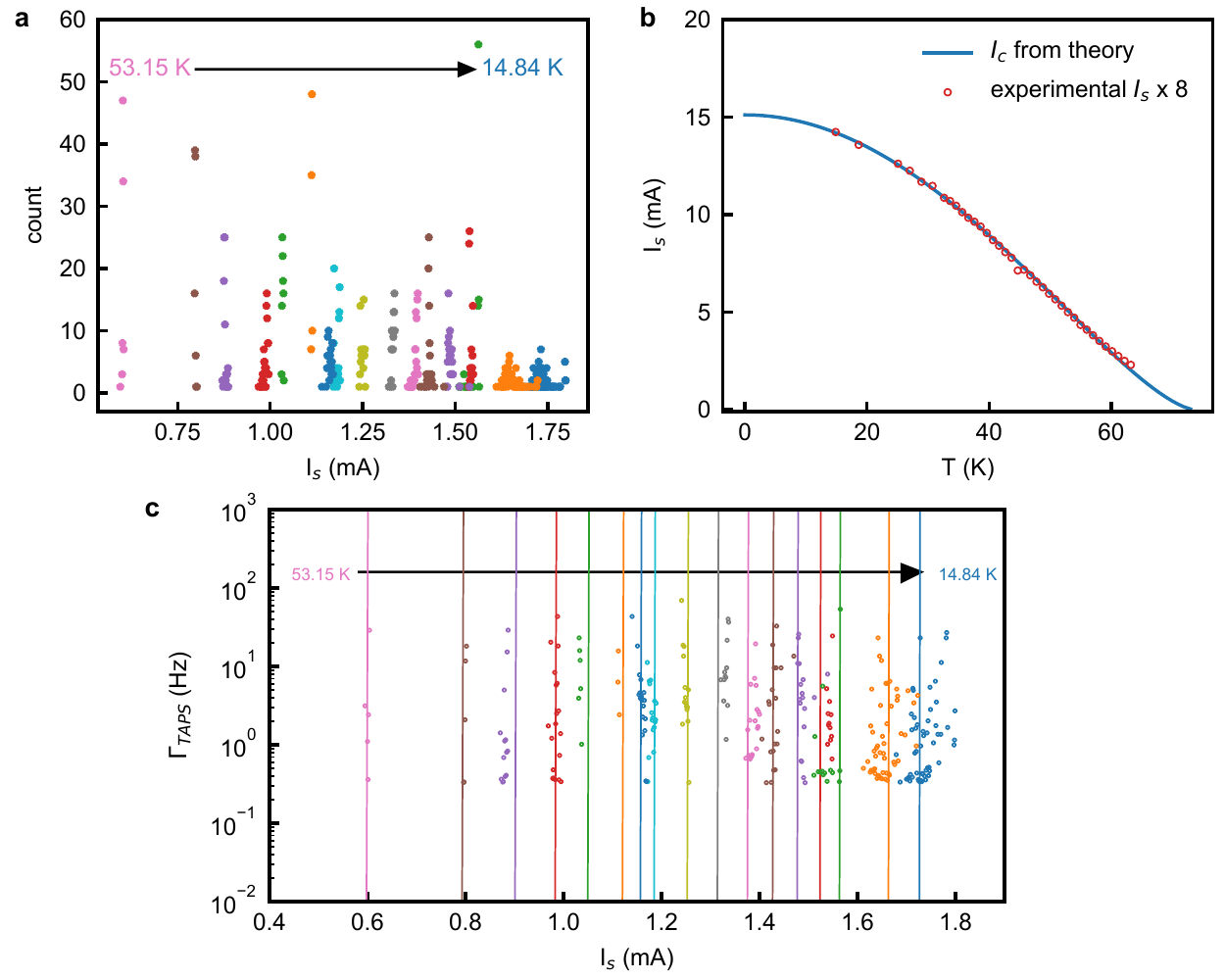}
\caption{ \label{fig:figS10}{\footnotesize (a) Switching statistics of NW2 at different temperatures. As the temperature is reduced the switching current distribution width increases \citep{sahu_individual_2009}. (b) Experimentally obtained switching current ($I_s$) and depairing current ($I_c$) from theory using (4). Switching current is 8 times smaller than the depairing current. (c) Switching rate from experiment (open circles) and TAPS rate from theory (solid lines). }}
\end{figure}

\section{Protocol to map photovoltage over nanowires}

\begin{figure}[!ht]
\includegraphics[width=7.6cm]{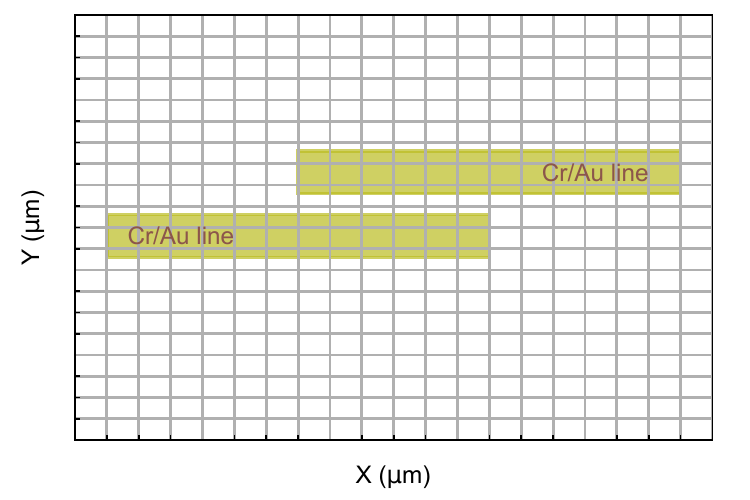}
\caption{ \label{fig:figS10}{\footnotesize we first identify the switching currents of the nanowires. The nanowire is then biased with a fixed $dc$ current near one of the identified switching currents. As a next step, we focus a laser beam on the nanowire. We divide the entire region of interest into smaller grids. We then move the laser spot to each square area and measure photovoltage across the nanowire. The relative movement between laser and sample is done by an X-Y piezo scanner on which the nanowire device is mounted. Collected photovoltages are then combined to get spatial mappings of photovoltage. }}
\end{figure}

\clearpage

%-----------------------------------------------------------------

\end{document}